# Thermal Conductivity of Metastable Ionic Liquid [C$_2$mim][CH$_3$SO$_3$]

D. Lozano-Martín[a,b], S.I.C. Vieira[a,‡], X. Paredes[a], M.J.V. Lourenço[a], C.A. Nieto de Castro[a,*], J. V. Sengers[c] and K. Massonne[d]

[a] Centro de Química Estrutural, Faculdade de Ciências, Universidade de Lisboa, Campo Grande, 1749-016 Lisboa, Portugal

[b] Grupo de Termodinámica y Calibración (TERMOCAL), Research Institute on Bioeconomy, Escuela de Ingenierías Industriales, Universidad de Valladolid, Paseo del Cauce, 59, E-47011 Valladolid, Spain

[c] Institute for Physical Science and Technology, University of Maryland, College Park, MD 20742, USA

[d] BASF SE, RC/OI - M300, 67056 Ludwigshafen, Germany

*Corresponding author. E-Mail: cacastro@ciencias.ulisboa.pt

‡Present address: CERENA-Centre for Natural Resources and the Environment, Instituto Superior Técnico, Av. Rovisco Pais, 1049-001 Lisboa, Portugal

**Abstract**

Ionic liquids have been suggested as new engineering fluids, namely in the area of heat transfer, as alternatives to current biphenyl and diphenyl oxide, alkylated aromatics and dimethyl polysiloxane oils, which degrade above 200 °C and pose some environmental problems. Recently, we have proposed 1-ethyl-3-methylimidazolium methanesulfonate, [$C_2$mim][$CH_3SO_3$], as a new heat transfer fluid, because of its thermophysical and toxicological properties. However, there were some interesting points raised in this work, namely the possibility of the existence of liquid metastability below the melting point (303 K) or second order-disorder transitions ($\lambda$-type) before reaching the calorimetric freezing point.

This paper analyses in more detail this zone of the phase diagram of the pure fluid, by reporting accurate thermal-conductivity measurements between 278 and 355 K with an estimated uncertainty of 2 % at a 95% confidence level. A new value of the melting temperature is also reported, $T_{melt}$ = 307.8 ± 1 K.

Results obtained support liquid metastability behaviour in the solid-phase region and permit the use of this ionic liquid at a heat transfer fluid at temperatures below its melting point.

Thermal conductivity models based on Bridgman theory and estimation formulas were also used in this work, failing to predict the experimental data within its uncertainty.

**Introduction**

The ionic liquid 1-ethyl-3-methylimidazolium methanesulfonate, [$C_2$mim][$CH_3SO_3$] (Trade marks Basionics® ST35; ECOENG™ 110), is a fluid with several industrial applications, but also with a very interesting structural behaviour. In a research project with BASF, we were asked to measure its thermophysical properties and evaluate its capacity to be used as a new heat transfer fluid, including density, speed of sound, heat capacity, viscosity, electrical, and thermal conductivity of the industrial product, as it arrived at the laboratory and after drying under vacuum, for the temperature range 283.15–363.15 K, at atmospheric pressure [1]. The measurements of thermal conductivity and heat capacity showed an interesting phase behaviour below the melting point (303.8 K) [2], which could suggest the existence of second order–disorder transitions ($\lambda$-type) or liquid metastability.

As discussed in our previous publication, many authors, including us [1 and references therein], have reported values of property measurements, like density, speed of sound, viscosity, electrical conductivity and heat capacity for temperatures, down to 273.15 K, well below 303.8 K, without noticing any deviations from "liquid-like" behaviour. We also reported a rheometric study of this ionic liquid at 310 K, above the quoted melting-point, and found that it does not show a totally Newtonian behaviour, having a residual shear stress at very low shear rates, necessary to be applied before the liquid starts to flow, typical of a Bingham fluid [1]. We noticed that, for shear rates greater than 4 $s^{-1}$, it could be assumed that the fluid behaves like a Newtonian fluid, which is the case in many applications.

For heat capacity, we could only perform measurements with DSC above the melting temperature, with an expanded ($k$=2) relative uncertainty of 5%, and although a sudden increase of this property was found when approaching it, suggesting a possible second-order phase transition ($\lambda$-type) between two liquid phases, these results require further confirmation with a higher accuracy calorimeter.

For the case of thermal conductivity, measurements were performed from the higher to the lower temperatures – cooling, passing the reported melting temperature, all data points staying in a straight line for both liquid and supercooled liquid phases down to 263.15 K (-10°C). We stop cooling the sample, let it warm overnight and reheat to 304.80 K, where a maximum thermal conductivity of 0.301 $W·m^{-1}·K^{-1}$ was obtained, and then to 309.15 K, where the "liquid" behaviour was obtained again, 0.205 W/m/K. We then concluded that we could have an aspect of an order-disorder phenomena or metastable phases, which should be explored further.

This paper analyses in more detail this zone of the phase diagram of the pure fluid, by reporting accurate thermal conductivity measurements between 278 and 355 K with an estimated uncertainty of 2 %, at a 95% confidence level.

**Results**

Extensive measurements of the thermal conductivity of ST35 were performed. The transient hot-wire probe instrument described below, permitted the measurement with several heat power dissipations applied (and consequently different temperature rises), proving that there was no indication of convection effects. Measurements were made with different cycles in temperature. In addition, two samples of Basionics® ST35, furnished by BASF, were prepared.

**Sample 1** - pure ST35: obtained by mixing a solid sample extracted from the blue-barrel of BASF at room temperature, plus an old liquid sample, investigated previously [1]. The sample was mixed mechanically with a glass rod several times before and during the measurements, remained in a solid + liquid (sludge) state, maybe due to the mechanical agitation and heating by friction. The initial water content was 24220 ± 430 ppm. See Figure 1A for the solid taken from the barrel.

The results for the thermal conductivity $\lambda$ as a function of the reference temperature $T_{ref}$ are depicted below for sample 1 in Figure 2 and reported in Table S1. $T_{ref}$ is defined as the average between the initial probe temperature and the final probe temperature of each measurement, as is normal in thermal-conductivity measurements performed with the transient hot-wire method [3]. Measurements were obtained at several days, between 31/10/2019 and 20/11/2019. The measurements started at $T_{ref} \approx$ 297 K (24°C) in the "sludge" state, increasing the temperature until $T_{ref} \approx$ 355 K (82°C), and then decreasing the temperature directly to 283 K (10°C) and 278 K (5°C). The sample was in the "sludge" state before $T_{ref} \approx$ (306-307) K (33 – 34) °C, became liquid until the maximum temperature (see Figure 1B), and then a sudden drop in the value of thermal conductivity was found. The temperature of the peak of the lambda shape (maximum thermal conductivity) is just above the melting point $T_{melt} \approx$ 303.8 K (30.6°C) for pure 1-ethyl-3-methylimidazolium methanesulfonate [2], and below the quoted melting point of ST35 by BASF 308 K (35°C). However, upon decreasing the temperature to 283 K (5°C), the sample remained liquid, confirming the results already published [1]. The water content of the sample after this set of measurements was measured by means of Karl-Fisher titration and found to be 12770 ± 390 ppm. It stayed liquid, for several months (now around room temperature, 295 K (22°C) until we dropped inside a small amount of new solid ST35 from the barrel, to trigger crystallization in the supercooled liquid, as used by Blesic et al. [2]. Nucleation started and sample turned solid (sludge), a strong evidence of breaking a metastable state – see Figure 1C.

**Sample 2** – pure ST35, solid: obtained directly from the blue barrel of BASF. Impossible to extract a sample to measure its moisture because it was solid before starting the measurements. The results for the thermal conductivity $\lambda$ as a function of the reference temperature $T_{ref}$ are depicted in Figure 3 for sample 2 (blue circles) when compared to those of the sample 1 results (black circles).

Only "heating" measurements were performed. Measurements started in $T_{ref} \approx$ 308 K (5°C), increasing the temperature until $T_{ref} \approx$ 320 K (47°. The sample started at the solid state and became liquid after the transition at $T_{ref} \approx$ (306-307) K (33 – 34) °C, but with fairly higher values of the thermal conductivity for the solid sample 2 against the " sludge" sample 1, at temperatures below that of the peak of the transition (solid has higher thermal conductivity than "sludge"). The values at the liquid state for sample 2 confirm that of the sample 1. The water content after the measurements was 12810 ± 200 ppm for sample 2, like that of sample 1 after the measurements.

It is clear from Figure 3, that the shape of the peak is better defined, as many experiments were made by taking small temperature steps (around 1 °C). It appears also, that the amplitude of the enhancement is larger in this case, where the maximum value of the thermal conductivity was found to be $\approx$ 0.40 W·m$^{-1}$·K$^{-1}$, compared with 0.33 W·m$^{-1}$·K$^{-1}$ for sample 1. We show in this figure also the unpublished data of Kreek et al. (2016) [4] (green open squares) and Bioucas et al. (2018) [1] (red crosses). The agreement with the new data is very good, especially for sample 1, where the temperature cycle was essentially the same. However, the transition in the thermal conductivity, in the heating "path" is very pronounced and consistently happening at a temperature near 307.8 K (34.6°C), four degrees above the reported melting point [2]. It is interesting that BASF always considered this ionic liquid as an undercooled melt and reported the melting point of ST35 at 35°C, and therefore the registered name of the industrial compound (ST35).

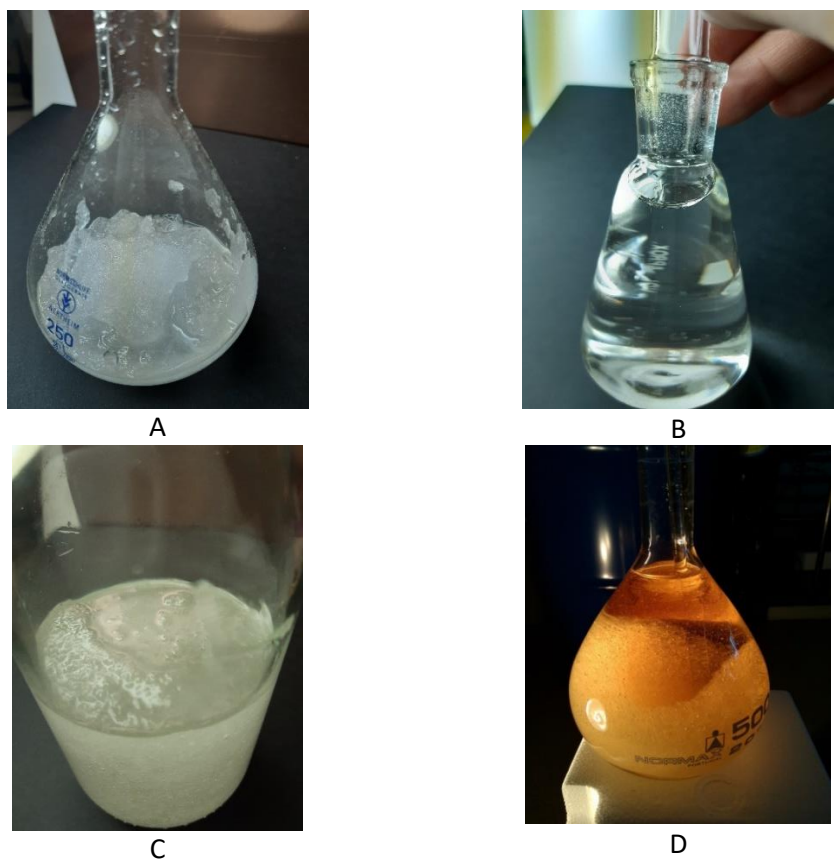

Figure 1 – Sample pictures. Picture A – Solid ST35 taken from blue-barrel; Picture B – Liquid; Picture C – Solid/Liquid ST35, nucleated sample 1; Picture D – 10 K below the expected melting point (303.8 K) [1], under yellow light. (Picture D Reprinted with permission from Bioucas, F. E. B.; Vieira, S. I. C.; Lourenço, M. J. V.; Santos, F. J. V.; Nieto de Castro, C. A.; Massonne, K. [C$_2$mim][CH$_3$SO$_3$] – A Suitable New Heat Transfer Fluid? Part 1. Thermophysical and Toxicological Properties. Ind. Eng. Chem. Res., 2018, 57 (25), 8541–8551. Copyright (2018) American Chemical Society).

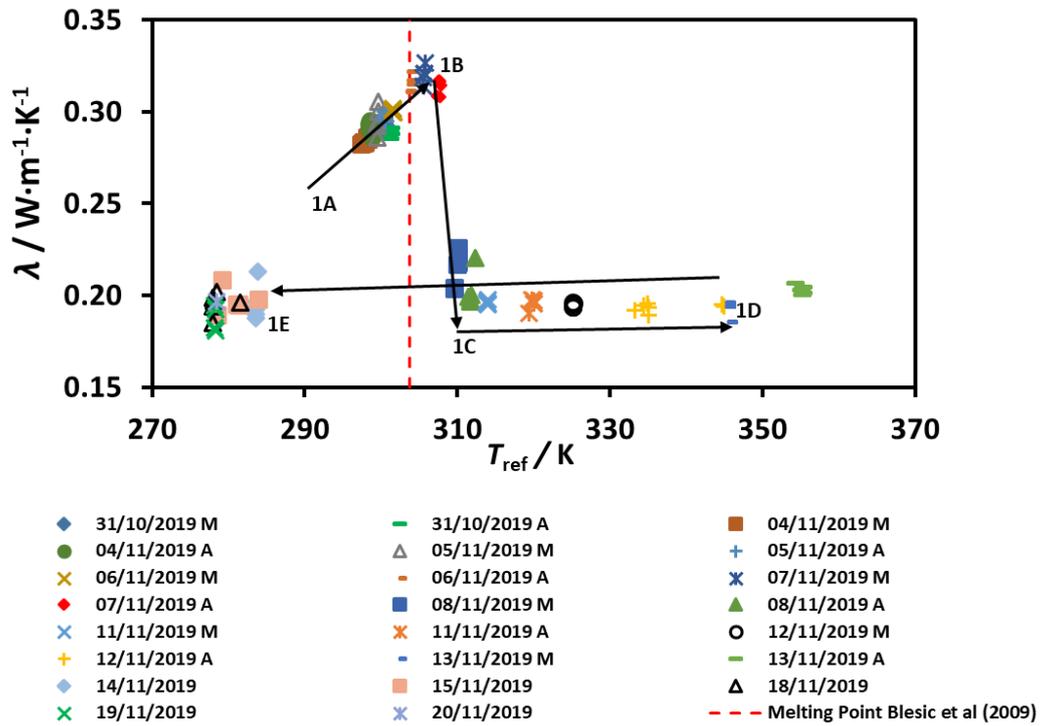

Figure 2 – The thermal conductivity of [C$_2$mim][CH$_3$SO$_3$] – sample 1. The legend indicates the dates with "M" = morning, and "A" = afternoon. Arrows indicate the temperature variation, 1A to 1D (heating) and 1D to 1E (cooling).

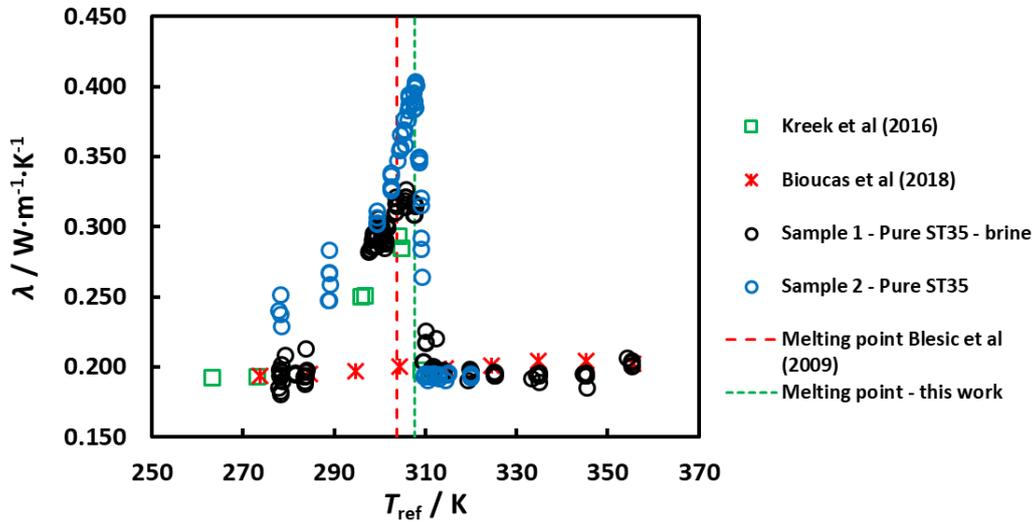

Figure 3 – The thermal conductivity of [C$_2$mim][CH$_3$SO$_3$] – sample 2, 5 – 47 °C ("heating", starting from solid). Also shown data for Sample 1 and published data of Bioucas et al. (2018) [1] and unpublished data taken in our laboratory by Kreek et al. (2016) [4]. All data agree with a melting temperature near 307.8 ± 1 K (34.6 ± 1) °C.

P a g e  6 | 15

**Discussion**

Our experiments reveal a remarkable behaviour of the thermal conductivity of [$C_2$mim][$CH_3SO_3$] which strongly depends on whether the measurements are taken upon increasing or decreasing the temperature of the sample. In principle we could consider a number of scenarios, namely:

a) The presence of a second-order continuous phase transition between two solid states according the nomenclature of Ehrenfest [5] and recently discussed by Jaeger [6]. In addition, a glass transition in the compound could be possible.
b) The presence of a liquid-liquid phase separation with a critical point of mixing as discussed by Kurita et al. [7] and which also occurs in supercooled water [8].
c) The presence of a normal first-order phase transition between the liquid and solid state with a melting temperature of 307.8 K (34.6°C).
d) The presence of a liquid metastable state in the supercooled liquid below the melting temperature down to low temperatures, around 283 K (5°C).

It should be pointed out that the thermal conductivity can only diverge at a second-order phase transition, where the mass density is the order parameter as is the case near the vapour-liquid critical point, but not when the concentration is the order parameter [9]. For instance, in a phase separating mixture with a critical point of mixing the mass conductivity diverges, but not the thermal conductivity [9,10]. Neither does the thermal conductivity diverge upon approaching the liquid-liquid critical point in supercooled water, where the entropy is the order parameter [8,11]. So, our observed increase of the thermal conductivity is not compatible with either a) or b). Moreover, it has been demonstrated from DSC measurements that a glass transition in [$C_2$mim][$CH_3SO_3$] occurs at 201.196 K [12] or 211 K [13] and predictive methods have indicated its occurrence at 210.3 K [14] and 206.92 K [15], temperatures that are much lower than the lowest temperature (278 K) in our experiments. In addition, a demixing liquid-liquid phase transition is unlikely, also because the refractive index would be different in the two phases. However, Seki et al. [16] did not find any deviation of the refractive index from a linear temperature dependence down to 283.15 K, a result confirmed by Freire et al. [17] down to 288.15 K.

Hence, our experiments are only consistent with hypotheses c) and d). First, there is definitely a first-order melting phase transition at 307.8 K (34.6 °C) as demonstrated by the sudden drop of the thermal conductivity as a function of temperature when increasing the temperature starting from the solid phase to the liquid phase (Fig. 3), observed for all the samples investigated by us, as well as by Kreek et al. [4] and by Bioucas et al. [1].

Blesic et al. [2] have reported a melting temperature as $T_{melt}$ = (303.75±3) K which is lower than our value $T_{melt}$ = 307.8 K. Blesic et al. [2] deduced the melting temperature of [$C_2$mim][$CH_3SO_3$] from DSC, upon heating with a scanning rate 5 K·min$^{-1}$, using an in-house made sample without a determination of its purity or water content. In their paper the dependency of the melting point upon the cationic chain (Fig. 4) shows a very large decrease from C1 (91.0) to C2 (30.6) and then an increase to C3 (39.8) and C4 (73.7), which is very strange. There might be an unaccounted error in the reported value or in the uncertainty statement.

In addition, their paper was not collected in the ILThermo Database [18], and no other melting-point data was found, except the value reported in databases in the literature [19-22], industrial and commercial suppliers MSDS's, and reported by BASF [23] as 35 °C (308 K), and therefore the registered name of the industrial compound (ST35). We propose the new value for $T_{melt}$ = 307.8 ± 1 K.

Hypothesis d) is proved by the current work, i.e., there is a liquid metastable state present in the supercooled liquid, below the melting point, that goes to low temperatures (around 5°C), when the temperature is lowered from above the melting point. This agrees also with the reported data for viscosity, heat capacity, density, speed of sound (see analysis in [1]) electrical conductivity and refractive index [16,17], where there is no change in slope in the property variation with temperature, when entering the supercooled liquid.

A similar behaviour to that encountered for thermal conductivity near the melting line has been found by Vila et al. [24], for the electrical conductivity of several ionic liquids with the same cation (1-ethyl-3-methyl imidazolium, $C_2mim^+$), when performing measurements for increasing and decreasing temperatures, passing the solid-liquid or liquid-solid transitions. Data showed the existence of hysteresis loops for some ILs, while others did not present any transition (at least not measurable), without any jump in the electrical conductivity. Several anions were studied ($Cl^-$, $Br^-$, $I^-$, $BF_4^-$, $PF_6^-$ and $CH_3-C_6H_6-SO_3^-$) with different water contents (up to 5000 ppm, dealer certification). As mentioned by these authors, the existence of hysteresis loops is associated to underlying phase transitions in which abrupt changes of some involve physical quantity take place, as well as absorption or release of energy as latent heats. And they do not appear in many other ILs, like [$C_2mim$][$C_2H_5SO_4$] [25]. That hysteresis loop appears when temperature decreases below $T_{melt}$, but the electrical conductivity of the IL decreased with temperature is the same than that in its liquid state for some degrees (from 10 K to 60 K depending on the IL). However, waiting some time can produce a sudden drop on the electrical conductivity, meaning solidification of the sample. The authors also reported that the form and existence of the hysteresis loops depends on the crystallinity degree of each compound.

Although no results are available for the electrical conductivity of [$C_2mim$][$CH_3SO_3$] at temperatures below 293 K [1], it is conceived that the electrical conductivity will follow the liquid like behaviour until 263 K or below and that adding a solidification nucleus will drop to very small values of the solid state magnitude.

The effect of the amount of water in the properties of this IL will be the subject of a future publication [26]. As the amount of water in the samples studied is high (~ 1.3 %, 12800 ppm), the effect of the small molecules of water in the structure of the ionic liquid could influence the melting transition and the metastability of our sample. However, the effect of water on the thermal conductivity of the IL amounts to less than 1% for $x_{water}$ < 0.27 ($w_{water}$ < 0.97) [26], so any effect in the thermal conductivity of our samples was not detectable, as it is much smaller than the experimental uncertainty.

To estimate the dependence of the thermal conductivity on temperature we consider a formula from Bridgman [27]:

$$\lambda = 2.8 k_B (v^*)^{-2/3} u = 2.8 \, k_B (v^*)^{-2/3} \left(\frac{1}{\rho \kappa_S}\right)^{1/2} \qquad (1)$$

where $\lambda$ is the thermal conductivity of the liquid, $k_B$ is the Boltzmann constant, $v^*$ is the molecular volume (equal to the ratio of the molecular mass divided by the mass density), $u$ the speed of sound, in the limit of zero frequency, $\rho$ the density and $\kappa_S$ the adiabatic compressibility. As this quantity is very difficult to obtain, we can use Eyring modification [28], as described by Biddle et al. [10]:

$$\lambda = 2.8 k_B (v^*)^{-2/3} \left(\frac{1}{\rho \kappa_T}\right)^{1/2} \qquad (2)$$

where $\kappa_T$ the isothermal compressibility. This quantity is related to the speed of sound $u$ through the relation:

$$u = \left(\frac{C_P}{C_V} \frac{1}{\rho \kappa_T}\right)^{1/2} \qquad (3)$$

And therefore, Eyring expression can be transformed to:

$$\lambda = 2.8 k_B (v^*)^{-2/3} \gamma^{-1/2} u \qquad (4)$$

Unfortunately there are no data for either $\gamma$ or $\kappa_T$ for [$C_2$mim][$CH_3SO_3$], but we can estimate its value from Eq. (3), by using data for a similar ionic liquid [$C_2$mim][$C_2H_5SO_4$] obtained by Nieto de Castro et al. [29]. This value is 1.30 at 273.15 K and 1.27 at 343.15 K.

In Fig. 4 we have plotted the experimental data for the thermal conductivity of the ionic liquid, earlier obtained by Bioucas et al [1] (orange symbols) and those obtained in the present work (open symbols), the actual experimental data, corrected to water-free values at nominal temperatures, as described in the section on Materials and Methods, are listed in Table 1. They can be represented by a linear function of the temperature $T$:

$$\lambda(\text{W·m}^{-1}\text{·s}^{-1}) = a_1 + a_2 T(\text{K}) \qquad (6)$$

For the data of Bioucas et al. [1], the coefficients in this correlation have the values $a_1$ = 0.15868 W·m$^{-1}$·K$^{-1}$, $a_2$ = 1.300E-04 W·m$^{-1}$·K$^{-2}$, with a root mean square deviation, at a 95% confidence level, of 0.0031 W·m$^{-1}$·K$^{-1}$. No point deviates by more than 1.2 % (the highest temperature one), which is commensurate with the expanded ($k$ = 2) uncertainty of the data, 2% [1]. For the case of the present data, the correlation has $a_1$ = 0.17523 W·m$^{-1}$·K$^{-1}$ and $a_2$ = 6.260E-05 W·m$^{-1}$·K$^{-2}$, with a root mean square deviation, at a 95% confidence level, of 0.0048 W·m$^{-1}$·K$^{-1}$. The maximum deviation of experimental points is never larger than 1.3% except at for the highest temperature, 2.7%, also commensurate with the uncertainty of the experimental measurements described in the next section. Both correlations agree with each other within 1 % at the lower temperature (273.15 K) and 3.7 % at the highest temperature (358.15 K).

We applied Eyring modification of Bridgman's model by using Eq. (3) for our polyatomic ionic liquid, with $\gamma$ obtained from Eq. (3) for [$C_2$mim][$C_2H_5SO_4$] from [29], and the values of the necessary properties for [$C_2$mim][$CH_3SO_3$] at 0.1 MPa taken from [1]. It can be seen that the calculated value for the polyatomic liquid is around 30% lower than the experiment, with a negative variation with temperature, contrary to the positive slope found experimentally. In addition, Bridgman model generates a higher thermal conductivity than Eyring model, which is rather unusual, a fact already found by Bridgman in its original paper [27].

**Table 1. Thermal conductivity of [C$_2$mim][CH$_3$SO$_3$] for the liquid and metastable liquid states**

| Current work | | Bioucas et al [1] | |
|---|---|---|---|
| $T_{nom}$ / K | $\lambda_{wf}(T_{nom})$ /W·m$^{-1}$·K$^{-1}$ | $T_{nom}$ / K | $\lambda_{wf}(T_{nom})$ /W·m$^{-1}$·K$^{-1}$ |
| 280 | 0.1954 | 275 | 0.1944 |
| 283 | 0.1928 | 285 | 0.1957 |
| 310 | 0.1933 | 295 | 0.1970 |
| 312 | 0.1941 | 305 | 0.1983 |
| 314 | 0.1961 | 315 | 0.1996 |
| 315 | 0.1939 | 325 | 0.2009 |
| 320 | 0.1964 | 335 | 0.2022 |
| 320 | 0.1935 | 345 | 0.2035 |
| 325 | 0.1948 | 355 | 0.2048 |
| 335 | 0.1939 | | |
| 345 | 0.1943 | | |
| 355 | 0.2028 | | |

Expanded relative uncertainty $U_r(\lambda)$ = 2 %, at a 95 % confidence level ($k$=2); $U_r$ = 0.02 K

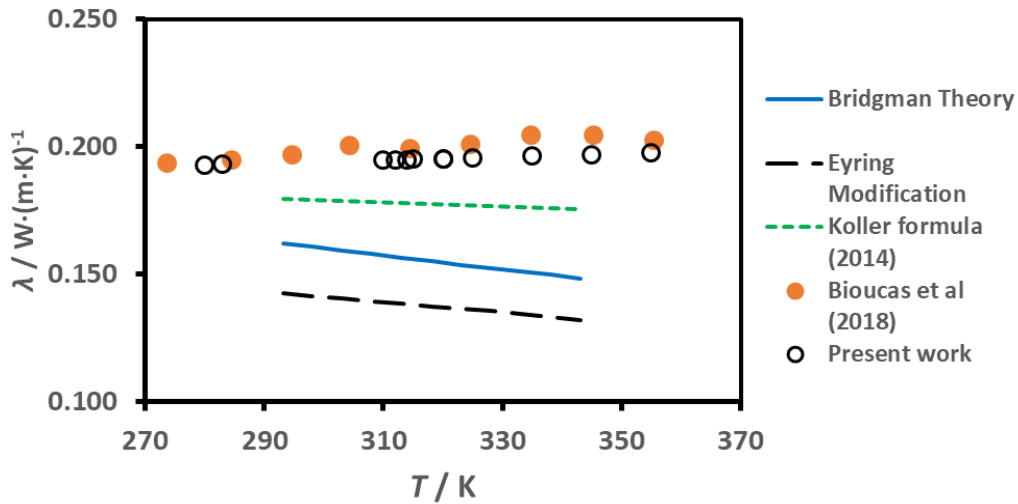

Figure 4 – The thermal conductivity of [C$_2$mim][CH$_3$SO$_3$] as a function of temperature.

There are several empirical methods to estimate the thermal conductivity of ionic liquids, namely those of Koller et al. [30], based on similarly behaviour of density and thermal conductivity, and others based on group contribution methods, namely those derived by Gardas and Coutinho [31] and Wu et al. [32]. These last two were developed using earlier data on thermal conductivity [31,32] and using estimated values of the critical temperature of ionic liquids [32], obtained with high uncertainty due to the small vapour pressure of ionic liquids, which decompose before reaching this gas-liquid critical point. We decide then to compare our experimental data with Kohler formula [30], and the results are also shown in Fig.4. This empirical method underestimates also the thermal conductivity of [C$_2$mim][CH$_3$SO$_3$] by 8% at 293 K and 14% at 343 K, a very interesting result as this compound was not used in the development of the estimation formula.

**Materials and Methods**

The [$C_2$mim][$CH_3SO_3$] used in this study (CAS Number 145022-45-3), was obtained from BASF, under the trade name of Basionics® ST35, with an assay ≥ 97% with ≤ 0.5% water and chloride ($Cl^-$) ≤ 2% [23]. The ST35 MSDS shows this ionic liquid must be handled in accordance with good industrial hygiene and safety practices, that wearing of closed work clothing is required additionally to the stated personal protection equipment and contact with the skin, eyes and clothing must be avoided. The water content of [$C_2$mim][$CH_3SO_3$] was always measured with Coulometric Karl Fischer titration (Metrohm 831) and the mass was measured using a Kern AEJ scale with an accuracy of $1 \times 10^{-5}$ g. Although thermal conductivity is not very sensitive to water content, like viscosity or electrical conductivity [1], all the experimental data obtained was corrected to water-free thermal conductivity data, by using the value of $\partial\lambda/\partial w_w$ at low concentrations of water, where $w_w$ is the water mass fraction in the sample (0.0128), from measurements on the thermal conductivity of [$C_2$mim][$CH_3SO_3$] + water mixtures [26]. This coefficient was found temperature independent, and equal to $0.2420 \cdot w_w$ W·m$^{-1}$·K$^{-1}$. These values are also displayed in Table 1, for the liquid and metastable liquid ranges.

Thermal conductivity, $\lambda$, was measured using commercial equipment (Hukseflux Thermal Sensors B.V., model TPSY02). This model uses a single-needle sensor TP08 (needle length of 70 mm with the junction at around 17 mm from the tip, and a diameter of 1.2 mm), with an accuracy quoted by the manufacturers of ± 0.02 W·m$^{-1}$·K$^{-1}$ and a temperature accuracy of ± 0.02 K. To ensure that the sensor is vertically kept at a stable temperature in a specially made stainless-steel vessel built with an approximated volume of 100 cm$^3$, previously described by Bioucas et al. [1]. The heating /cooling is done with a copper jacket connected to an oil bath. The uncertainty of the instrument was checked by measuring the thermal conductivity of MilliQ water between 283.6 and 344.8 K and comparing with the IUPAC recommended standard values [33]. Deviations from SRD correlation were smaller than 0.8%, a very good sign of excellent operation. The expanded global relative uncertainty of the data was found to be $U_r$ = 2% ($k$=2). For the temperatures above 323.15 K the probe had to be removed and cleaned and the sample had to be equilibrated between measurements for longer times to reduce convection heat currents.

In order to detect possible onset of convection near the transition temperatures of ST35, induced by metastability, measurements were done in water (low viscosity) around room temperature, with three different power inputs in the THW probe (LOW, MEDIUM and HIGH), giving temperature rises between 0.25 K and 0.70 K. No curvature in the straight-line fits $\Delta T_{rise}$ vs ln $t$, were found [4]. Using normal procedure for further accessing the measurements, Table 2 shows the results obtained and their correction to the nominal temperature of 295.15 K, using the value of $\partial\lambda/\partial T$ obtained from the IUPAC correlation [33] and Figure 5 the mean values for the LOW, MEDIUM and HIGH powers, their standard deviation, the average of all 12 points and the standard deviation of this average. It can be easily seen that there is no power dependence of the data, and that the values of the thermal conductivity agree within their mutual statistical standard deviations. These results also support that there is no convection in these measurements and that the working model for the transient hot-wire is valid [4]. It is expected that, a fortiori, no convection exists in the measurements of the thermal conductivity of [$C_2$mim][$CH_3SO_3$] herein reported, as it has viscosities in the temperature range studied one hundred times bigger than water. Same type of heat inputs and measuring times were used. In addition, the big enhancement in the thermal conductivity happens in the solid phase, and in the "sludge", where the metastable liquid state was broken.

**Table 2 - The thermal conductivity of water at the nominal temperature of 295.15 K – Example of measurements performed with different heat inputs**

| $T_{ref}$ | $\lambda$ | $\sigma$ | $\Delta T_{rise}$ | Heating | $10^3 \partial\lambda/\partial T$ | $T_{nom}$ | $\lambda(T_{nom})$ |
|---|---|---|---|---|---|---|---|
| K | W·m⁻¹·K⁻¹ | W·m⁻¹·K⁻¹ | K | q | W·m⁻¹·K⁻² | K | W·m⁻¹·K⁻¹ |
| 293.93 | 0.5946 | 0.0095 | 0.263 | Low | 1.815 | 295.15 | 0.5968 |
| 293.91 | 0.5979 | 0.0071 | 0.247 | Low | 1.815 | | 0.6001 |
| 293.89 | 0.5957 | 0.0081 | 0.263 | Low | 1.815 | | 0.5980 |
| 293.96 | 0.5927 | 0.0088 | 0.296 | Low | 1.814 | | 0.5948 |
| 295.84 | 0.6068 | 0.0051 | 0.476 | Medium | 1.772 | | 0.6056 |
| 295.75 | 0.6029 | 0.0063 | 0.525 | Medium | 1.774 | | 0.6018 |
| 295.83 | 0.5964 | 0.0059 | 0.525 | Medium | 1.772 | | 0.5952 |
| 297.87 | 0.5995 | 0.0029 | 0.688 | High | 1.726 | | 0.5948 |
| 297.83 | 0.6043 | 0.0033 | 0.688 | High | 1.727 | | 0.5997 |
| 297.75 | 0.6016 | 0.0046 | 0.688 | High | 1.729 | | 0.5971 |
| 297.89 | 0.6062 | 0.0040 | 0.688 | High | 1.726 | | 0.6015 |
| 297.82 | 0.6016 | 0.0033 | 0.705 | High | 1.728 | | 0.5970 |
| | | | | | | Average | 0.5985 |
| | | | | | | STD | 0.0033 |
| | | | | | | % | 0.55 |

[a] Expanded relative uncertainty $U(\lambda)$ = 2 %, at a 95 % confidence level ($k$=2)

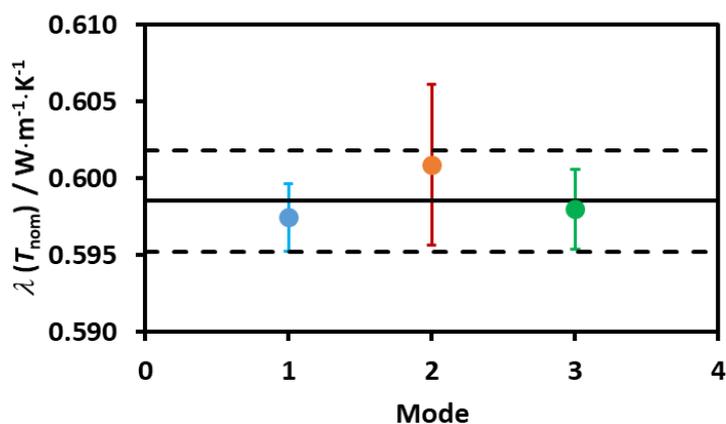

Figure 5 - The thermal conductivity of water at the nominal temperature of 295.15 K, measured with different heat powers. Mode 1 – LOW; Mode 2 – MEDIUM and Mode 3 – HIGH. Error bars for each set of measurements are show as the average of all data points (line) and its standard deviation (dashed line).

**Conclusions**

Measurements of the thermal conductivity of [C₂mim][CH₃SO₃] above and below the melting point show the existence of a metastable ionic liquid, which is very stable, and allows the measurement of several thermophysical properties typical of a "liquid" behaviour, like viscosity, electrical conductivity, speed of sound, heat capacity and refractive index. The thermal conductivity data obtained permit the determination of the melting temperature, with an uncertainty smaller than previous reported data. The thermal conductivity liquid model of Bridgman shows that it can predict its value qualitatively, 30% below the experiment, but with a different temperature dependence. Empirical estimation techniques were also analysed. All the measurements of the thermophysical properties of this compound so far reported by us and by other authors below the melting point, refer to this metastable liquid.

It is believed that this metastable behaviour can be safely used in several industrial applications, extending its temperature range to temperatures lower than the melting temperature.

**Acknowledgments**

The authors would like to thank Kristiina Kreek, PhD student of the University of Tallinn, with a training period of 4 months at our laboratory, during the spring of 2015, and contributing to some data for figure 3.

The authors would also like to thank BASF for supplying Basionics® ST35 and the funding agency FCT - Fundação para a Ciência e para a Tecnologia, Portugal, by its support through grant PEST OE/QUI/UI100/2013 and UID/QUIM/0100/2019 and UIDB/QUIM/0100/2020. Daniel Lozano-Martín thanks the University of Valladolid for its mobility grant.

**Author Contributions**

SICV made the first experiments with ST35 and detected the phenomena with MJVL, who extracted the samples from the blue-barrel, directed the measurements and selected the methodologies for calibrations, probe handling and temperature cycles. DLM and XP prepared the current samples and made the experimental measurements of the thermal conductivity. CANC wrote the manuscript, applied Bridgman theory to the data and participated with JVS in the theoretical interpretation of the observed metastability. KM reported the strange behaviour of this undercooled melt, providing the samples of ST35 and discussed the results.

**Conflicts of Interest**

There are no conflicts of interest. Information disclosed permitted by BASF.

**References**

1. Bioucas, F. E. B.; Vieira, S. I. C.; Lourenço, M. J. V.; Santos, F. J. V.; Nieto de Castro, C. A.; Massonne, K. [$C_2$mim][$CH_3SO_3$] – A Suitable New Heat Transfer Fluid? Part 1. Thermophysical and Toxicological Properties. *Ind. Eng. Chem. Res.*, **2018**, 57 (25), 8541–8551. DOI: https://doi.org/10.1021/acs.iecr.8b00723 .
2. Blesic, M.; Swadźba-Kwaśny, M.; Belhocine, T.; Nimal Gunaratne, H. Q.; Canongia Lopes, J. N.; Gomes, M.FR.C.; Pádua, A. A. H.; Seddon, K. R.; Rebelo, L. P. N. 1-Alkyl-3-methylimidazolium alkanesulfonate ionic liquids; [$C_nH_{2n+1}$mim][$C_kH_{2k+1}SO_3$]: synthesis and physicochemical properties. *Phys. Chem. Chem. Phys.*, **2009**, 11, 8939–8948. DOI: https://doi.org/10.1039/B910177M .
3. Nieto de Castro, C.A.; Lourenço, M.J.V. Towards the Correct Measurement of Thermal Conductivity of Ionic Melts and Nanofluids. Energies 2020, 13 (1), 99. Special Issue "Selected papers of the "1st International Conference on Nanofluids (ICNf). DOI: https://doi.org/10.3390/en13010099 .

4. Kreek, K.; Vieira, S.I.C.; Lourenço, M.J.V.; Nieto de Castro, C.A.; Koel, M. IoNanofluids from Graphene: A New Challenge. Poster presented at EUCHEM2016, Vienna, Austria, July 2-8, 2016, PP-108. http://euchem2016.book-of-abstracts.com/programme/confirmed-poster-presentations/ .
5. Ehrenfest, P. "Phasenumwandlungen im ueblichen und erweiterten Sinn, classifiziert nach dem entsprechenden Singularitaeten des thermodynamischen Potentiales." Verhandlingen der Koninklijke Akademie van Wetenschappen (Amsterdam) 36: 153–157; Communications from the Physical Laboratory of the University of Leiden, Supplement No. 75b (1933).
6. Jaeger, G. The Ehrenfest Classification of Phase Transitions: Introduction and Evolution. Arch. Hist. Exact Sci. 1998, 53, 51–81. DOI: https://doi.org/10.1007%2Fs004070050021 .
7. Kurita, R.; Murata, K.-I.; Tanaka, H. Control of fluidity and miscibility of a binary liquid mixture by the liquid–liquid transition. Nature Materials, 2008, 7, 647-652. DOI: https://doi.org/10.1038/nmat2225 .
8. Fuentevilla, D.A. and Anisimov, M. A. Scaled equation of state for supercooled water near the liquid-liquid critical point. Phys. Rev. Lett. 2006, 97, 195702. DOI: https://doi.org/10.1103/PhysRevLett.97.195702 ; ibid, 2007, 98, 149904. DOI: https://doi.org/10.1103/PhysRevLett.98.149904 .
9. Sengers, J.V. Transport properties of fluids near critical points. Int. J. Thermophys. 1985, 6, 203-232. DOI: https://doi.org/10.1007/BF00522145
10. Biddle, J.W.; Holten, V.; Sengers, J.V.; Anisimov, M.A. Thermal conductivity of supercooled water. Phys. Rev. E, 2013, 87, 042302. DOI: https://doi.org/10.1103/PhysRevE.87.042302 .
11. Sengers, J.V., in: Critical Phenomena, Proceedings of the International School of Physics "Enrico Fermi", Green M.S. ed. Academic Pres: New York, United States; 1971, pp. 445-567. ISBN: 0-12-368851-5
12. Seki S.; Kobayashi T.; Kobayashi Yo.; KatsuhitoT.; Miyashiro, H.; Hayamizu, K.; Tsuzukib, S.; Mitsugi, T.; Umebayashi, Y. Effect of cation and anion on physical properties of room temperature ionic liquids. J. Mol. Liq. 2010, 152, 9-13. DOI: https://doi.org/10.1016/j.molliq.2009.10.008 .
13. Gardas, R.L.; Costa, H.F.; Freire, M.G.; Carvalho, P.J.; Marrucho, I.M.; Fonseca, I.M.A., Ferreira, A.G.M., Coutinho, J.A.P. Densities and derived thermodynamic properties of imidazolium-,pyridinium-, pyrrolidinium-, and piperidinium-based ionic liquids. J. Chem. Eng. Data. 2008, 53, 805-811. DOI: https://doi.org/10.1021/je700670k .
14. Harris K.R.; Kanakubo, M. Self-diffusion coefficients and related transport properties for a number of fragile ionic liquids. J Chem Eng Data. 2016, 61, 2399-2411. DOI: https://doi.org/10.1021/acs.jced.6b00021 .
15. Safarov, J.; Huseynova, G.; Bashirov, M.; Hassel, E.; Abdulagatov, I. M. Viscosity of 1-ethyl-3-methylimidazolium methanesulfonate over a wide range of temperature and Vogel–Tamman–Fulcher model. Physics and Chemistry of Liquids 2018, 56:6, 703-717. DOI: https://doi.org/10.1080/00319104.2017.1379080 .
16. Seki, S.; Tsuzuki, S.; Hayamizu, K.; Umebayashi, Y.; Serizawa, N.; Takei, K.; Miyashiro, H. Comprehensive Refractive Index Property for Room-Temperature Ionic Liquids. J. Chem. Eng. Data, 2012, 57(8), 2211-2216. DOI: https://doi.org/10.1021/je201289w .
17. Freire, M. G.; Teles, A. R. R.; Rocha, M. A. A.; Schroder, B.; Neves, C. M. S. S.; Carvalho, P. J.; Evtuguin, D. V.; Santos, L. M. N. B. F.; Coutinho, J. A. P. Thermophysical Characterization of Ionic Liquids Able To Dissolve Biomass. J. Chem. Eng. Data, 2011, 56(12), 4813-4822. DOI: https://doi.org/10.1021/je200790q
18. Kazakov, A.; Magee, J.W.; Chirico, R.D.; Paulechka, E.; Diky, V.; Muzny, C.D.; Kroenlein, K.; Frenkel, M. NIST Standard Reference Database 147: NIST Ionic Liquids Database - (ILThermo),

Version 2.0, National Institute of Standards and Technology, Gaithersburg MD, 20899, http://ilthermo.boulder.nist.gov . Updated to June 11 2019.
19. ChemicalBook. 2017. 1-ETHYL-3-METHYLIMIDAZOLIUM METHANESULFONATE. [online]. Available at:
https://www.chemicalbook.com/ChemicalProductProperty_EN_CB6113535.htm
20. Molbase Encyclopedia. 2020. 1-Ethyl-3-Methylimidazolium Methanesulfonate. [online] Available at: http://www.chemspider.com/Chemical-Structure.17339776.html
21. Chemspider.com. 2020. 1-Ethyl-3-Methyl-1H-Imidazol-3-Ium Methanesulfonate | C7H14N2O3S | Chemspider. [online].
Available at: http://www.chemspider.com/Chemical-Structure.17339776.html
22. ChemSrc. 2020. 1-Ethyl-3-Methylimidazolium Methanesulfonate. [online] Available at: https://www.chemsrc.com/en/cas/145022-45-3_27072.html
23. BASF Safety data sheet according to Regulation (EC) No. 1907/2006. Revised: 05.09.2016 Version: 10.0.
24. Vila, J.; Fernández-Castro, B.; Rilo, E.; Carrete, J.; Domínguez-Pérez, M.; Rodríguez, J.R.; García, M.; Varela, L.M.; Cabeza, O. Liquid–solid–liquid phase transition hysteresis loops in the ionic conductivity of ten imidazolium-based ionic liquids. Fluid Phase Eq. 2012, 320, 1–10. DOI: https://doi.org/10.1016/j.fluid.2012.02.006 .
25. Vila, J.; Franjo, C.; Pico, J.M.; Varela, L.M.; Cabeza, O. Temperature Behavior of the Electrical Conductivity of Emim-Based Ionic Liquids in Liquid and Solid States", Portugaliae Electrochimica Acta 2007, 25, 163-172. DOI: https://doi.org/10.4152/pea.200701163 .
26. Bioucas, F.E.B.; Vieira, S.I.C.; Paredes, X; Santos, A.F.; Lourenço, M.J.V.; Santos, F.J.V.; Lopes, M.L.M.; Nieto de Castro, C. A.; Massonne, K. [C$_2$mim][CH$_3$SO$_3$] – A suitable new heat transfer fluid? Part 2. Thermophysical Properties of Its Mixtures with Water, Ind. Eng. Chem. Res., 2020 (manuscript in preparation).
27. Bridgman, P.W., The Thermal Conductivity of liquids, P. W. Proc. Natl. Acad. Sci. USA, 1923, 9, 341-345. DOI: https://doi.org/10.1073/pnas.9.10.341 .
28. Hirschfelder, J.O.; Curtiss, C.F.; Bird, R.B. Molecular Theory of Gases and Liquids, Wiley, New York; Chapman & Hall, London, 1954, p. 634. ISBN: 978-0-471-40065-3.
29. Nieto de Castro C.A.; Langa, E.; Morais, A.L.; Matos Lopes, M.L.; Lourenço, M.J.V.; Santos, F.J.V.; Santos, M.S.C.C.S.; Canongia Lopes, J. N.; Veiga, H. I.M.; Macatrão, M; Esperança, J.M.S. S.; Rebelo, L.P.N.; Marques, C.S.; Afonso, C.A.M., Studies on the density, heat capacity, surface tension and infinite dilution diffusion with the ionic liquids [C4mim][NTf2]; [C4mim][dca]; [C2mim][EtOSO3] and [aliquat][dca]. Fluid Phase Eq. 2010, 294, 157-179. DOI: https://doi.org/10.1016/j.fluid.2010.03.010 .
30. Koller, T.M.; Schmid, S.R.; Sachnov, S.J.; Rausch, M.H.; Wasserscheid, P.; Fröba, A.P. Measurement and Prediction of the Thermal Conductivity of Tricyanomethanide- and Tetracyanoborate-Based Imidazolium Ionic Liquids, Int. J. Thermophys., 2014, 35, 195-217. DOI: https://doi.org/10.1007/s10765-014-1617-1 .
31. Gardas, R.L.; Coutinho, J.A.P. Group contribution methods for the prediction of thermophysical and transport properties of ionic liquids, AIChE J., 2009, 55, 1274–1290. DOI: https://doi.org/10.1002/aic.11737 .
32. Wu, K-J.; Zhao, C-K; He, C-H. Development of a group contribution method for determination of thermal conductivity of ionic liquids, Fluid Phase Equilibria 339 (2013) 10–14. DOI: https://doi.org/10.1016/j.fluid.2012.11.024 .
33. Ramires, M.L.V.; Nieto de Castro, C.A.; Nagasaka, Y.; Nagashima, A.; Assael, M.J.; Wakeham, W.A., Standard Reference Data for the Thermal Conductivity of Water. J. Phys. Chem. Ref. Data. 1995, 24, 1377-1381. DOI: https://doi.org/10.1063/1.555963 .